\pdfoutput=1

\documentclass[a4paper,11pt]{article}
\usepackage{graphicx}
\usepackage{overpic}
\usepackage{verbatim} 
\usepackage[T1]{fontenc}

\usepackage{xspace}

\newcommand{\pT}{$p_{T}$\xspace} 

\newenvironment{centreFigure}
{
\begin{center}
\hspace*{-0.5in}\hspace*{-\marginparwidth}\begin{minipage}[b]{0.9\paperwidth}
\begin{center}
}
{
\end{center}
\end{minipage}
\end{center}
}

\begin{document}

\title{The HBOM Method for Unfolding Detector Effects.}

\author{J.~W.~Monk \& C.~Oropeza-Barrera
\\\
\\\
\begin{minipage}[c]{0.49\linewidth}
 Dept. of Physics \& Astronomy\\\
 University College London\\\
 Gower Street, London, England
\end{minipage}
\begin{minipage}[c]{0.49\linewidth}
\begin{flushright}
School of Physics \& Astronomy\\\
University of Glasgow\\\
Glasgow, Scotland
\end{flushright}
\end{minipage}
}
\maketitle

\begin{abstract}

We present the \emph{Hit Backspace Once More} (HBOM) method for correcting a measurement for the effect of an imperfect detector.  The HBOM method is a model-independent and potentially data-driven technique that repeatedly applies a parameterisation of the detector effect to observed data.  The correction is determined by extrapolating the data so-obtained to a detector effect of zero.  We demonstrate this technique using the two particle correlation function, which is an observable that can otherwise be difficult to correct for systematic shifts introduced by the detector.

\end{abstract}

\section{Introduction}

A measurement made on data from a collider experiment will normally include artefacts arising from the interaction of the final state particles with the detector.  Such ``detector effects'' include energy and angular smearing, production of additional particles through secondary scatterings, and inefficiencies in the detection of particles.  A measurement that contains detector artefacts is of less use than one that does not; in order to correctly compare such a result to either theoretical predictions or other measurements of the same observable, one needs access to a full description of the detector effects.  Detailed knowledge about the detector, typically in the form of a sophisticated simulation package, usually only exists within the experimental collaboration responsible.  There is no guarantee, or indeed expectation, that the detailed detector simulation will be available many years from now.

A good measurement will, therefore, de-convolute the detector effects from the fundamental physics processes and publish results that have been unfolded to match what one would observe with a ``perfect'' detector.  Such results are directly comparable to a hadron-level Monte Carlo (MC) simulation with no need for a detector simulation. 

Several methods for unfolding the detector effects exist.  One of the simplest is bin-by-bin unfolding, for which the measurement is performed on an ensemble of Monte Carlo simulations both with and without a detector simulation.  Correction factors are determined for each bin of the result from the ratio of the value with and without the detector simulation.  Bin-by-bin unfolding has the advantages that it is easy to understand and can work quite effectively on simple observables.  However, bin-by-bin unfolding does introduce a dependence on the Monte Carlo model(s) used to derive the bin-to-bin migration correction factors and, especially for more complicated observables, it tends to produce a bias in the shape of distributions.

Bayesian unfolding \cite{D'Agostini:1994zf}   expresses the probability that, given an observed value $X_{i}^{obs}$, the true value is $X_{j}^{true}$.   This expression requires knowledge of the set of probabilities that a true value $X_{j}^{true}$ will be observed as $X_{i}^{obs}$, together with a prior model for the probability distribution of the true values.  An iterative procedure is applied within Bayesian unfolding such that the corrected data from the first iteration is used as the prior for a second iteration.  

Iterative Bayesian unfolding has the advantages that it works even for complex observables and in principle, for a sufficiently large statistical sample, it contains negligible model dependence.  However, the iterative sequence may take a long time to converge, which can introduce large fluctuations into the result.  The choice of prior can also have an effect on the unfolded distributions if there are an insufficient number of iterations.    Bayesian unfolding is also quite an involved procedure and can be conceptually hard.  This latter point is not insignificant because in order to properly assign to a measurement any systematic uncertainties arising from the unfolding procedure, one needs to fully understand the process.  

Any unfolding procedure is simply trying to quantify as well as possible the effect that the detector has on an observable.  Conceptually, one may consider a detector to be a (very complicated) mathematical operator, $A$, that has been applied to the observable, $X_{i}$: $X_{i}^{obs}=X_{i}\left(0\right) = AX_{i}^{true} = AX_{i}\left(-1\right)$.  If $A$, or a reasonable approximation to it, is known then it may be applied \emph{again} to $X_{i}\left(0\right)$ to give $X_{i}\left(1\right) = AX_{i}\left(0\right) = AAX_{i}^{true}$.  In this way (and without knowing $X_{i}\left(-1\right)$) a sequence of identical observables with an increasingly large detector effect can be constructed: $X_{i}\left(0\right), X_{i}\left(1\right),\linebreak[1] X_{i}\left(2\right)...\linebreak[1]X_{i}\left(N\right)$.  Under the assumption that $X_{i}\left(N\right)$ is a smooth function of $N$, a numerical fit can be made to the sequence of $X_{i}$.  Such a fit can be evaluated at $N=-1$ to provide an estimate of the true value of the observable in the absence of any detector effects.  

In this note we show an example of this method applied to a two particle correlations observable measured in an ensemble of Monte Carlo generated events.  Detector effects can often express themselves in a complicated way for correlation observables because the loss of a single particle from the detector makes a contribution to the observable that depends on all of the remaining particles in the same collision event.  As such, bin-by-bin unfolding does not work to correct the two particle correlation function because it introduces a significant bias to the shape of the function.

\section{Observable Definition}

The two particle correlation function has been used by several experiments in order to investigate soft non-perturbative effects and particle production mechanisms \cite{Eggert:1974ek,Ansorge:1988fg,Alver:2007wy,Khachatryan:2010gv}. The definition used here was fully inclusive for all events containing more than two charged particles.  All charged particles whose transverse momentum and pseudo-rapidity ($p_{T}$ and $\eta$, respectively) satisfied $p_{T} > 100$~MeV and $\left|\eta\right| < 2.5$ were accepted into the analysis.  The correlation function was defined in terms of a foreground distribution, $F\left(\Delta t\right)$, and a background distribution, $B\left(\Delta t\right)$, where $\Delta t$ was the separation between a pair of particles in either $\eta$ ($\Delta\eta$) or azimuth ($\Delta\phi$).  As such, the two particle correlation function probed the correlations between pairs of particles as a function of their $\Delta\eta$ or $\Delta\phi$ separation.

The foreground distribution was obtained by determining the $\Delta t$ for each pair of charged particles within the same event  and filling a histogram with those $\Delta t$ values weighted by $2 / N_{ch}$, where $N_{ch}$ is the multiplicity of charged particles in each event.  The multiplicity-dependent weighting factor is a standard feature of two particle correlations that gives each event a weight according to $\left(N_{ch}-1\right)/N_{p}$, where $N_{p}$ is the number of charged particle pairs per event.  The factor of $\left(N_{ch}-1\right)$ is present because the strength of correlations between particles was historically found to be approximately proportional to $\left(N_{ch}-1\right)$ \cite{Eggert:1974ek}.

The background distribution used was similar to the foreground distribution, but pairs of particles were taken from independent events before determining the $\Delta t$ distribution.  Unlike the foreground distribution, no multiplicity dependent weighting factor was used for the background.  The background distribution was instead normalised to unit integral.  By obtaining the correlations between particles in independent events, the background distribution reveals the contribution to correlations that arises solely from acceptance cuts (or, in a real experiment, effects caused by the detector).

Having obtained the foreground and background distributions, the two particle correlation function, $R\left(\Delta t\right)$, was produced as given by equation \ref{eqn:R}

\begin{equation}
R\left(\Delta t\right)=\frac{F\left(\Delta t\right)}{B\left(\Delta t\right)} - \left< N_{ch} - 1\right>_{evts}\label{eqn:R}
\end{equation}
where $\left< N_{ch} - 1\right>_{evts}$ is the value of $\left(N_{ch}-1\right)$ averaged over the ensemble of all events.  The general shape of $R\left(\Delta\eta\right)$ and $R\left(\Delta\phi\right)$ is shown in figure \ref{fig:series} of section \ref{sec:application}.

\section{Monte Carlo Sample}

A sample of 50 million minimum bias Monte Carlo events was generated with tune ``4C'' of the Pythia 8 Monte Carlo generator \cite{Sjostrand:2007gs}.  Tune 4C was produced after a comparison of Pythia 8 to early results from the Large Hadron Collider, which motivated a somewhat reduced cross section for single and double diffraction as well as an improved description of the average transverse momentum per charged particle \cite{Corke:2010yf}.  Mixed samples were produced containing non-diffractive, single and double-diffractive proton-proton collision events at a centre of mass energy of 7~TeV.  Events were analysed using the Rivet \cite{Buckley:2010ar} Monte Carlo analysis framework.

\section{Pseudo-detector Simulation}

\begin{figure}
\begin{minipage}[c]{0.49\linewidth}
\begin{overpic}[width=\columnwidth]{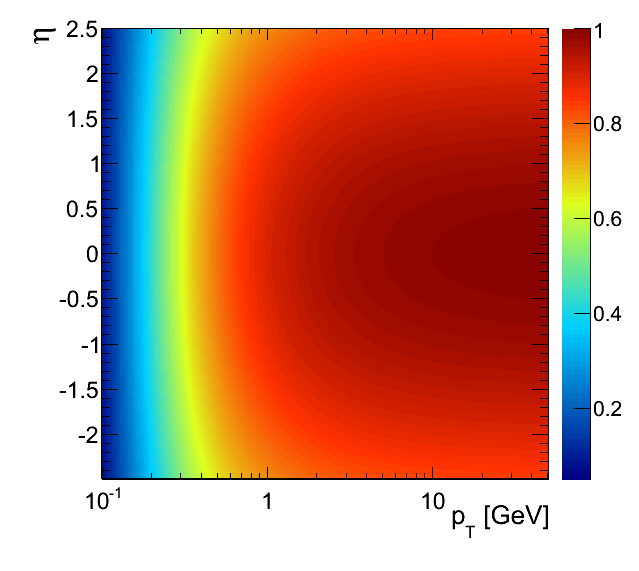}
\end{overpic}
\end{minipage}
\begin{minipage}[c]{0.49\linewidth}
\begin{overpic}[width=0.9\columnwidth]{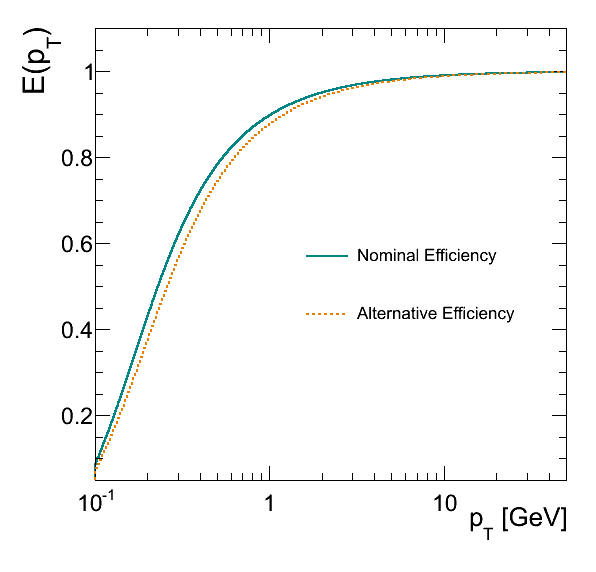}
\end{overpic}
\end{minipage}

\begin{minipage}[c]{0.49\linewidth}
\begin{overpic}[width=0.9\columnwidth]{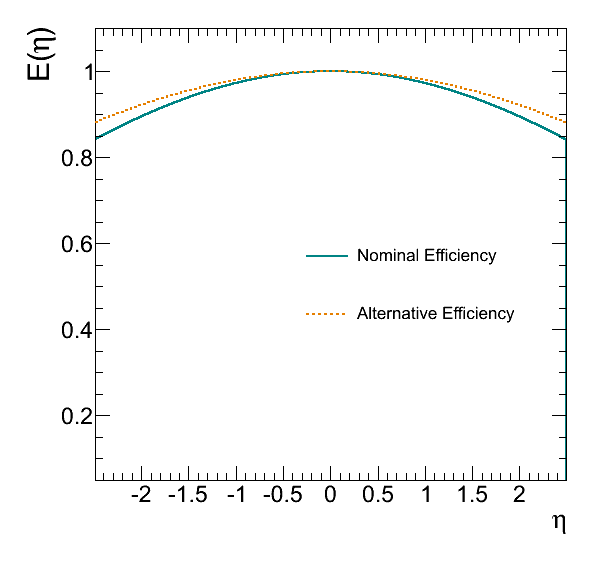}
\end{overpic}
\end{minipage}\hspace{0.02\linewidth}
\begin{minipage}[c]{0.45\linewidth}
\caption[Particle detection efficiency]{The particle detection efficiency as a function of $p_{T}$ and $\eta$.  Top left shows the 2D $p_{T}$ and $\eta$ dependence, while the top right and bottom left show the separate $p_{T}$ and $\eta$ dependence, respectively.  The nominal and alternative detector parameterisations are shown for the latter two plots.}\label{fig:detParam}
\end{minipage}
\end{figure}

The effect of a detector on the observed charged particles was parameterised as a function of a particle's \pT and  $\eta$.  A detector will typically have an efficiency for detecting charged particles that rises rapidly from around 50~MeV to nearly unit efficiency at around 1~GeV or more.  The efficiency for detecting a charged particle will typically fall gently as $\eta$ departs from zero.  The detector efficiency employed for this study was therefore in the form of an inverse tangent function of \pT and a broad Gaussian in $\eta$, as given in equation \ref{eqn:detParam}

\begin{equation}
E\left(p_{T}, \eta\right)=\frac{2}{\pi}\tan^{-1}\left(\frac{p_{T}-p_{T0}}{\sigma_{p}}\right)\exp\left(-\left(\frac{\eta}{\sigma_{\eta}}\right)^{2}\right)\label{eqn:detParam}
\end{equation}
The parameters $\sigma_{p}$, $p_{T0}$ and $\sigma_{\eta}$ determine how quickly the efficiency rises and falls with \pT and $\eta$ respectively.  Realistic nominal values of $\sigma_{p}=0.15$~GeV, $p_{T0}=80$~MeV and $\sigma_{\eta}=6$ were used, which resulted in the efficiency shown in figure \ref{fig:detParam}.  In order to investigate the effect of using a detector parameterisation that does not perfectly match the true detector effect (if, for example, the particle detection efficiency were not perfectly known or were mis-modelled) a second (henceforth the alternative) detector parameterisation was defined, for which values of $\sigma_{p}$ and $\sigma_{\eta}$ were taken as $0.18$~GeV and 7, respectively. 

The particle detection efficiency of equation \ref{eqn:detParam} was applied to the Monte Carlo sample by generating a unique random number in the range $\left\{0, 1\right\}$ for each particle in each event.  If the random number for a given particle was greater than its detection efficiency then that particle was removed from the sample.  This procedure was used for the initial generation of $X_{i}\left(0\right)$ from $X_{i}^{true}$ and for the subsequent generation of the sequence $X_{i}\left(1\right)...X_{i}\left(N\right)$.

In addition to employing an incorrect particle detection efficiency, the detector effect, $A$, that is repeatedly applied to $X_{i}\left(0\right)$ to obtain $X_{i}\left(1\right)...X_{i}\left(N\right)$ may lack certain aspects of the detector's behaviour.  In order to investigate this possibility, we also introduced additional particles into a sample of events to approximate the effect of secondary scatterings within the detector, but without including that effect in $A$.  The probability that a particle splits into a pair of particles was parameterised as a function of $\eta$ as in equation \ref{eqn:splitProb}

\begin{equation}
P\left(\eta\right)=\exp\left(\left(\frac{\eta}{12}\right)^{2}\right) - 0.97 \label{eqn:splitProb}
\end{equation}
The resulting pair of particles was given an opening angle $\theta_{s}$ with a probability $P\left(\theta_{s}\right)$ given in equation \ref{eqn:splitAngle}

\begin{equation}
P\left(\theta_{s}\right)=\sqrt{\frac{8}{\pi}}\left(1+\tan^{2}\left(\frac{\theta_{s}}{2}\right)\right)\exp\left(-8\tan^{2}\left(\frac{\theta_{s}}{2}\right)\right)\label{eqn:splitAngle}
\end{equation}
In general, a detector simulation should also include smearing of the particle energies and angles.  However, in this case, the observable did not have a strong dependence on particle energies because only a charged particle's $\eta$ and $\phi$ coordinate enter into the two-particle correlation function.  Thus energy smearing would only have a small effect due to the $p_{T}>100$~MeV charged particle cut, and as such was not considered here.  In any real experiment the observable measured here would be obtained from tracking detectors, which typically provide excellent angular resolutions of better than 0.1~milliradians.  Angular smearing was therefore also not considered, since any effect would be small in a realistic scenario.

\section{Application of the HBOM Method.}\label{sec:application}

The nominal detector parameterisation was applied to a sample of hadron-level truth events, from which the two particle correlation observable was then calculated.  The result is an approximation to what would be observed if the true events were detected in a real detector.  The same detector parameterisation was then applied a further five times to the same ensemble of events to give $X_{i}\left(1\right)...X_{i}\left(5\right)$. 

\begin{figure}[h]
\begin{centreFigure}
\begin{minipage}[c]{0.49\linewidth}
\begin{overpic}[width=\columnwidth]{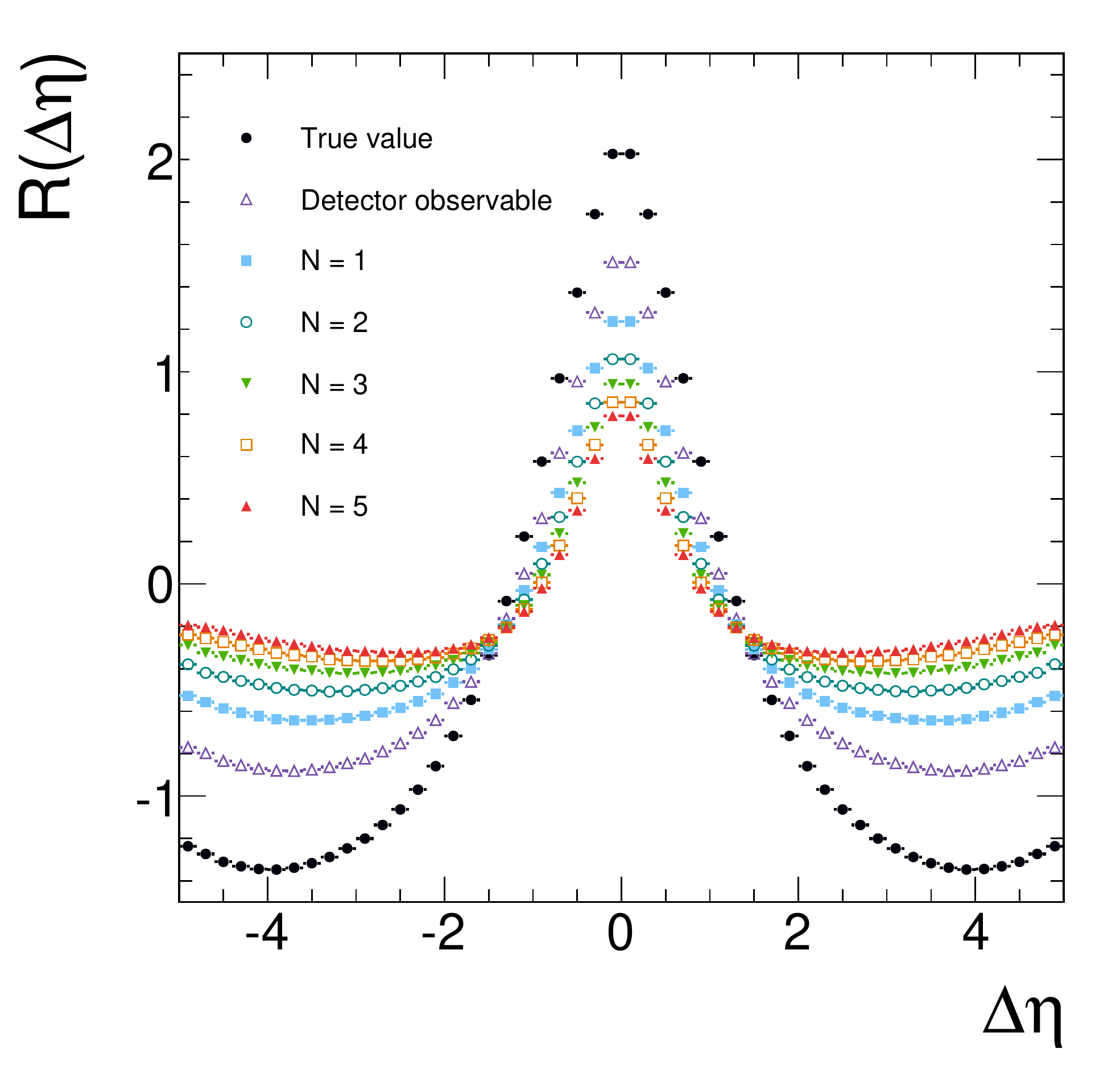}
\put(80,75){a)}
\end{overpic}
\end{minipage}
\begin{minipage}[c]{0.49\linewidth}
\begin{overpic}[width=\columnwidth]{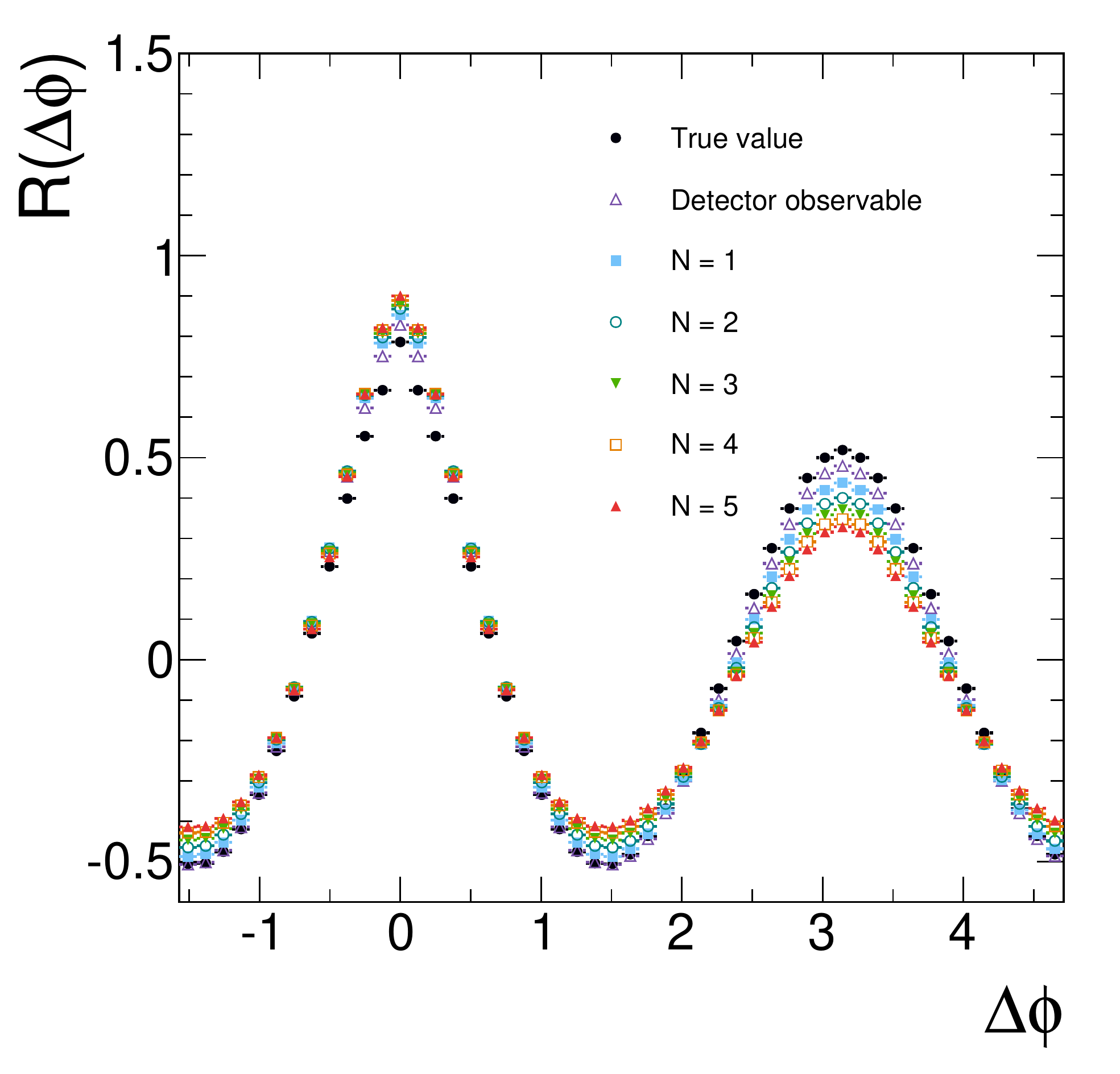}
\put(28,75){b)}
\end{overpic}
\end{minipage} 
\end{centreFigure}

\begin{center}
\caption[Series of increasingly large detector effects]{The $\Delta\eta$ (a) and $\Delta\phi$ (b) dependence of the two particle correlation observable.  The true value of the observable is shown as solid black circles, while the value observed by the parameterised nominal detector is shown as purple open triangles.  Further applications of the nominal detector to the observable show an increasing flattening of the correlation function in $\Delta\eta$ and a slight sharpening of the correlation in $\Delta\phi$ near zero.}\label{fig:series}
\end{center}

\end{figure}

The set of observables $X_{i}\left(0\right)...X_{i}\left(5\right)$ (each expressing an increasingly large detector effect) is shown together with the true value of the observable in figure \ref{fig:series}.  Note how the structure in the observable (the peak at $\Delta\eta=0$ and the gentle rise at high $\Delta\eta$) is flattened out with each additional application of the detector parameterisation.  Flattening occurred because, in removing particles from events, the detector effect weakened the correlations observed.  The amount of flattening decreases with each subsequent application of the detector.  In order to arrive at an estimate of the true observable, such flattening had to be reversed.  Conversely, the $\Delta\phi$ dependence of the two particle correlation function develops a somewhat sharper peak at $\Delta\phi=0$ (and correspondingly broader peak at $\Delta\phi=\pi$) as the detector effect is increased.  The different behaviour shown by the $\Delta\eta$ and $\Delta\phi$ dependence of the two particle correlation function illustrates why a simple re-scaling of the results tends not to work for complex observables.
 
\begin{figure}[h]
\begin{minipage}[c]{0.49\linewidth}
\begin{overpic}[width=\columnwidth]{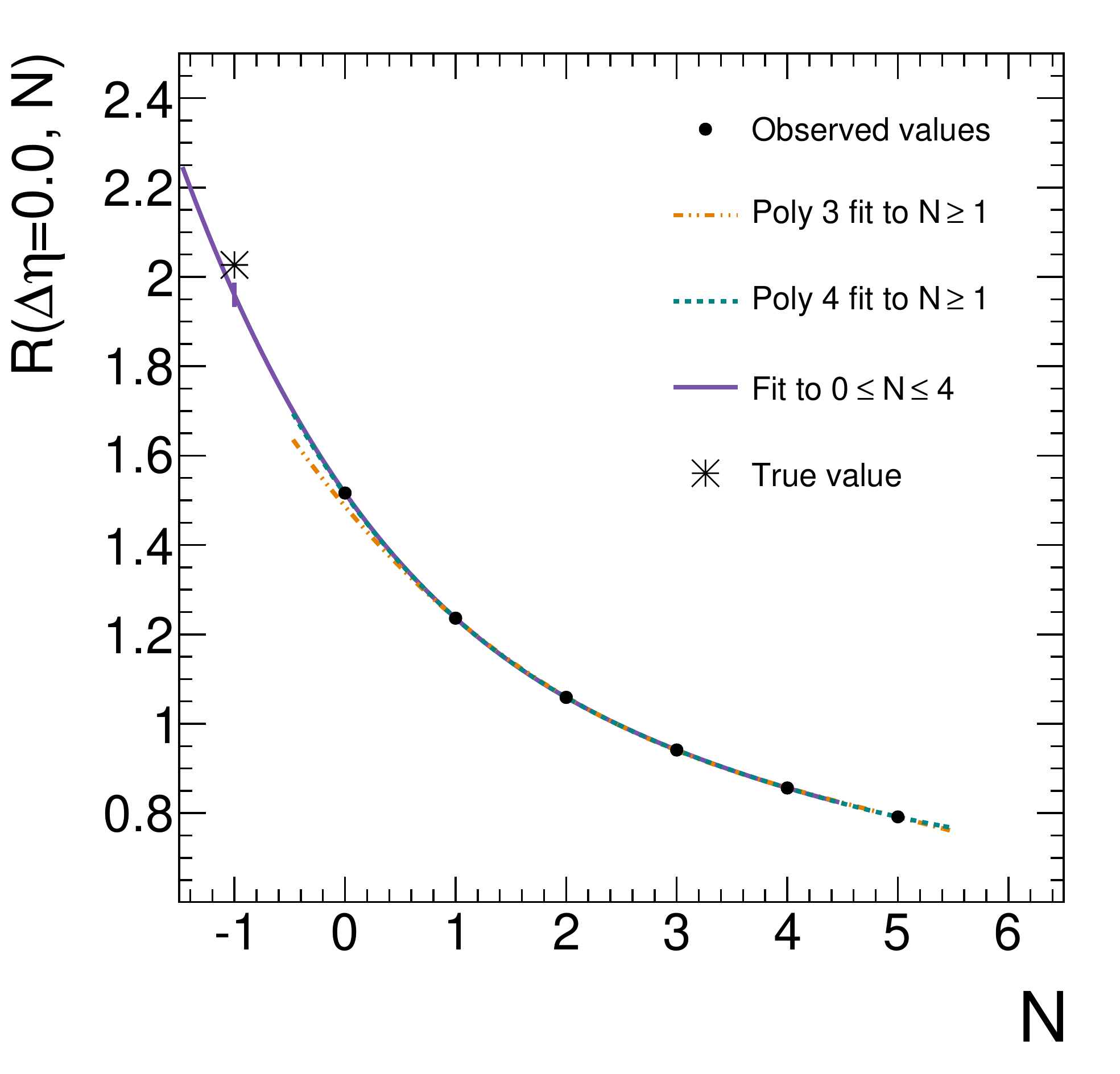}
\put(28,80){a)}
\end{overpic}
\end{minipage}
\begin{minipage}[c]{0.49\linewidth}
\begin{overpic}[width=\columnwidth]{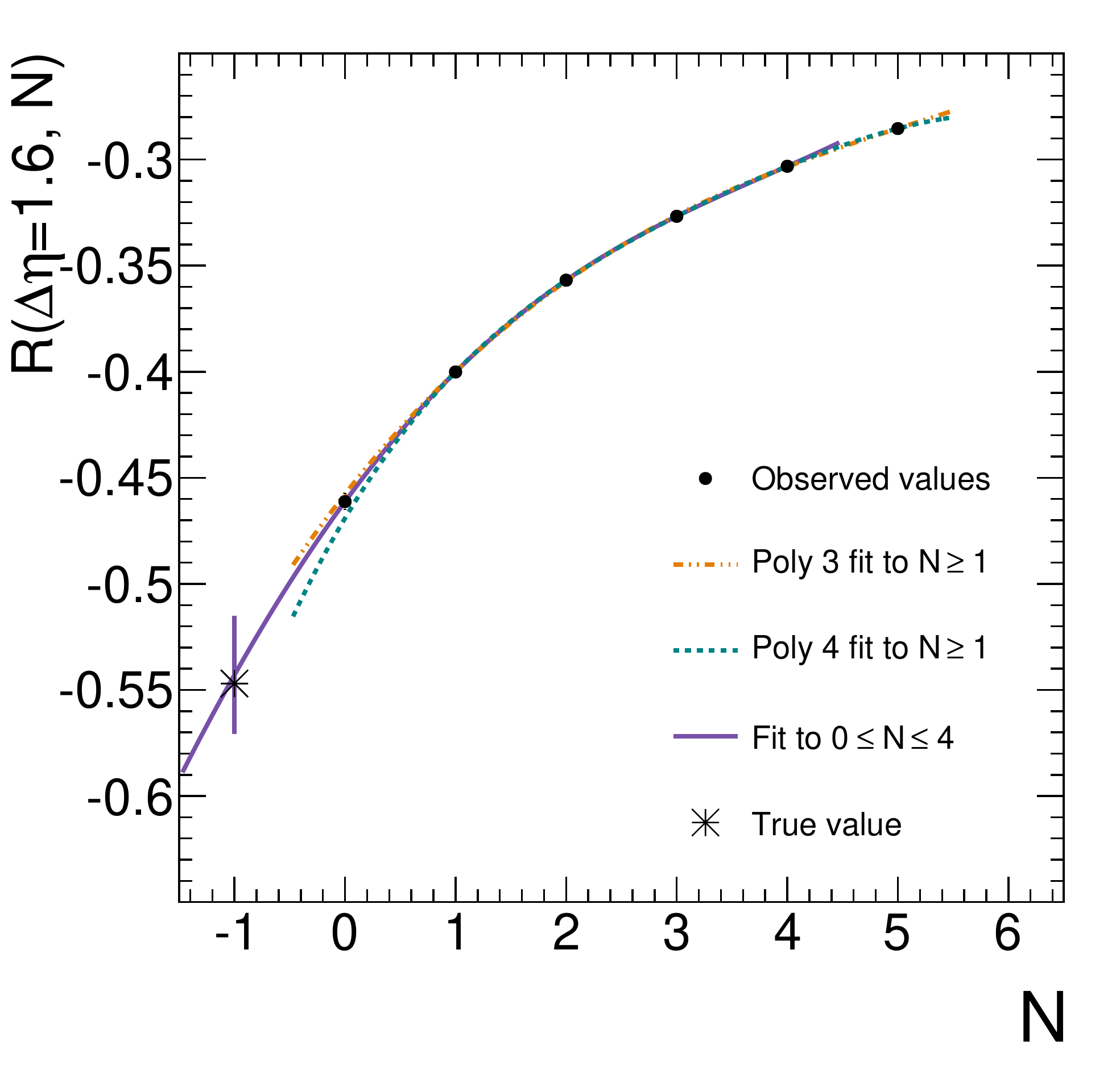}
\put(28,80){b)}
\end{overpic}
\end{minipage} 

\begin{minipage}[c]{0.49\linewidth}
\begin{overpic}[width=\columnwidth]{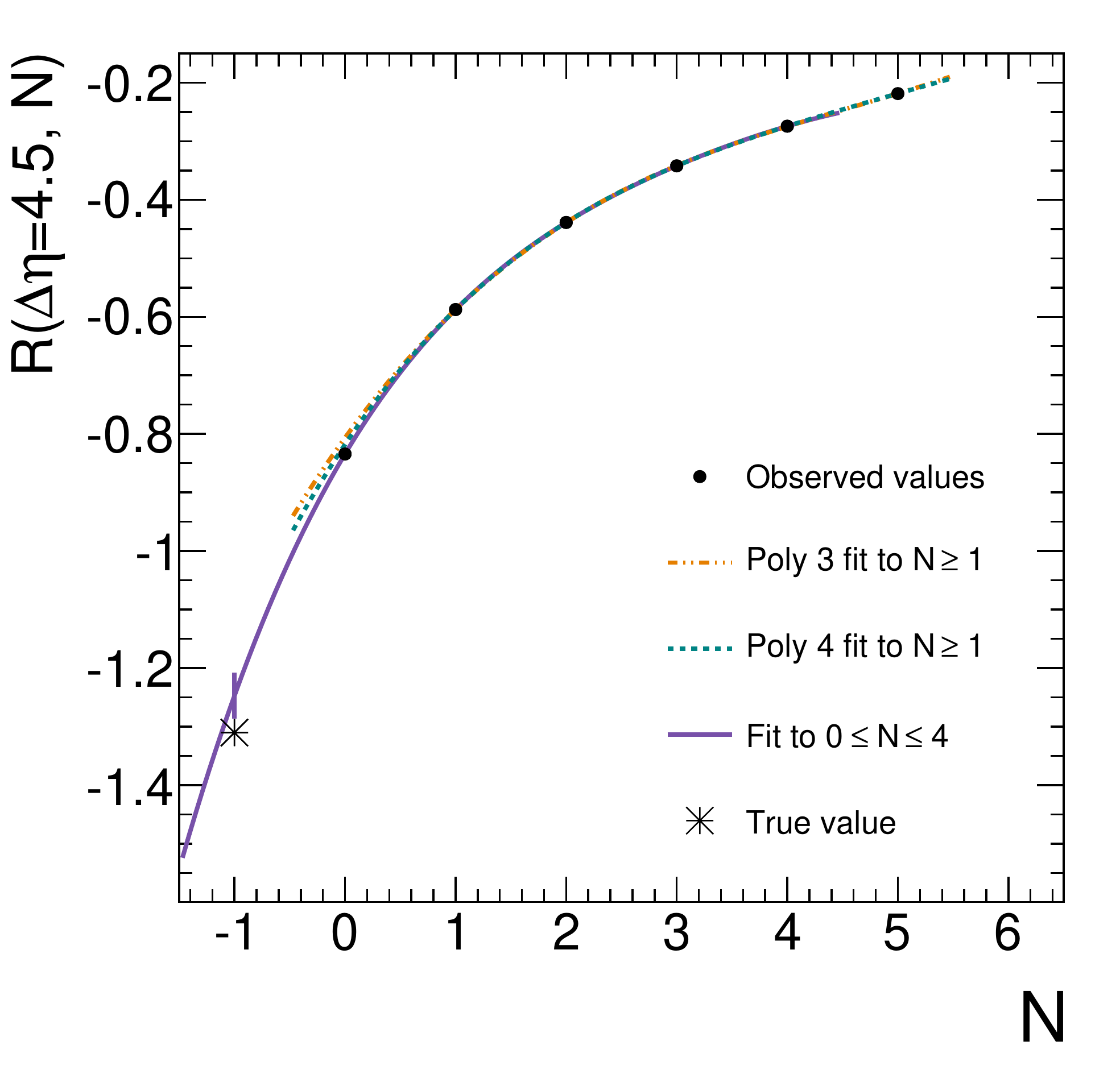}
\put(28,80){c)}
\end{overpic}
\end{minipage}\hspace*{0.05\linewidth}
\begin{minipage}[c]{0.53\linewidth}
\begin{flushleft}
\caption[Fits to detector iterations curves]{Polynomial fits to the value of the two particle correlation function Vs. the number of applications of the detector, $N$.  The value measured by the detector is at $N=0$, the true value is shown by a star at $N=-1$.   The observed $N=0$ value is estimated from both a cubic and quartic fit to the $N=1$ and higher points,  while the true $N=-1$ value is estimated from a quartic fit to the $N=0$ and higher points.  Fits are shown for three $\Delta\eta$ values of 0 (a), 1.6 (b) and 4.5 (c). }\label{fig:linearFits}
\end{flushleft}
\end{minipage} 

\end{figure}

The value of the observable in each of its bins was plotted as a function of the number of times the detector parameterisation had been applied.  A polynomial of degree four was fitted to the resulting ``iterations curve'' for each bin of the observable.  Figure \ref{fig:linearFits} shows the fits for three different points in the two particle correlation function at $\Delta\eta=0$, $\Delta\eta=1.6$ and $\Delta\eta=4.5$.  The values of $\Delta\eta$ shown here were chosen in order to be representative of the whole range of correlation function values.  Figure \ref{fig:linearFits} also shows additional fits that are discussed in section \ref{sec:fits}.  The true hadron-level value of the correlation function is shown in figure \ref{fig:linearFits} (but that point was not used in the fit), together with the estimate of that true value obtained from the fit and its corresponding fit uncertainty.  The agreement between the point estimated by the fit at $N=-1$ and the true value of the observable was found to be generally good.

\section{Results of the Correction Procedure}

\begin{figure}[h]
\begin{centreFigure}
\begin{minipage}[c]{0.49\linewidth}
\begin{overpic}[width=\columnwidth]{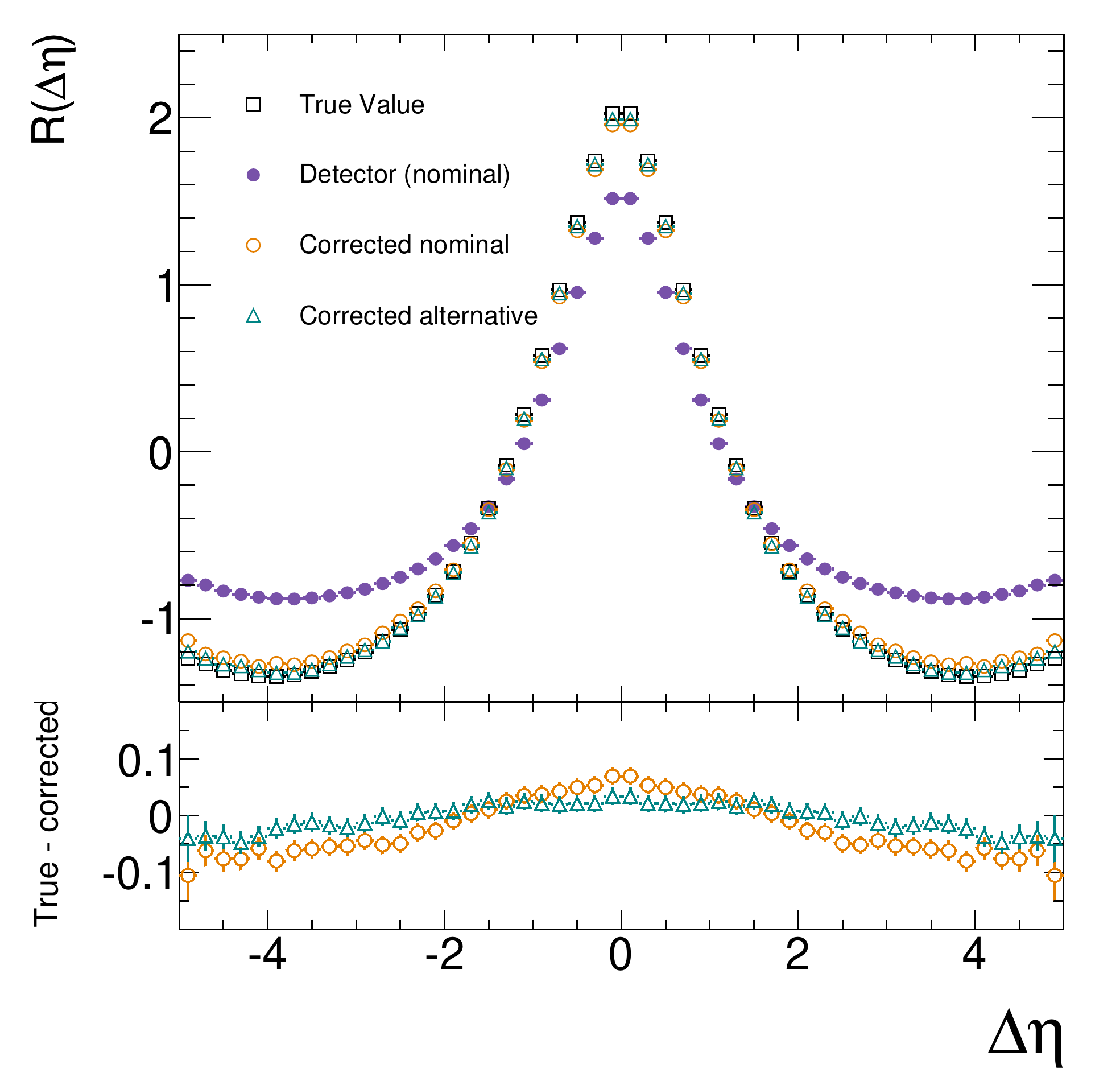}
\put(80,80){a)}
\end{overpic}
\end{minipage}
\begin{minipage}[c]{0.49\linewidth}
\begin{overpic}[width=\columnwidth]{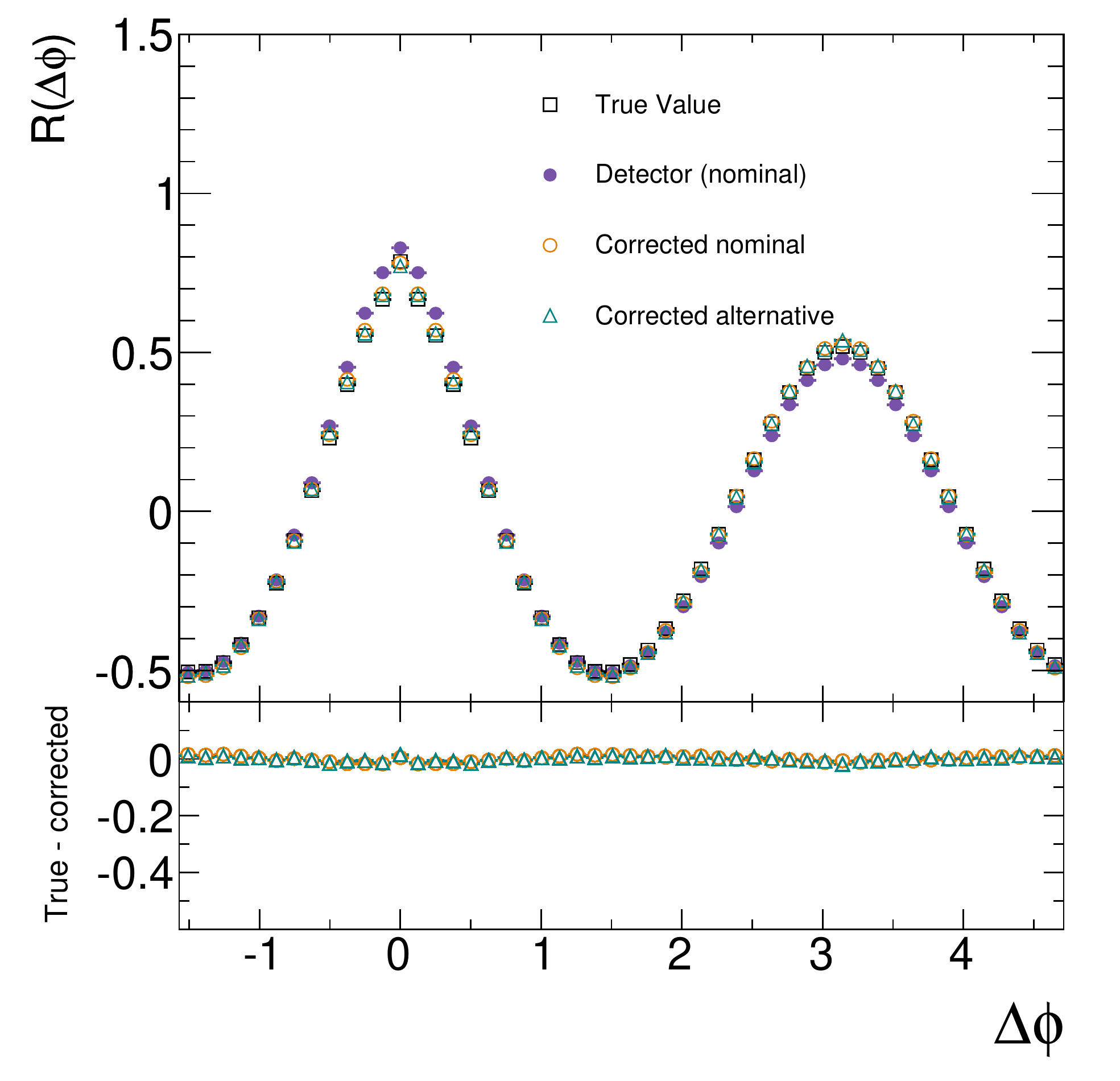}
\put(28,80){b)}
\end{overpic}
\end{minipage} 
\end{centreFigure}
\caption[Corrected and uncorrected two particle correlation function using a linear fit]{Corrected and uncorrected $\Delta\eta$ (a) and $\Delta\phi$ (b) dependence of the two particle correlation function.  Corrections were applied using linear polynomial fits of degree four to the iterations curves (see figure \ref{fig:linearFits}). The nominal detector-level observable is shown as filled circles (colour online).  The nominal detector after corrections is shown as un-filled circles, while a correction that uses the miss-modelled alternative detector is shown as un-filled triangles.  The true value is shown as un-filled squares and the differences between the true and corrected values are shown in the bottom panel.}\label{fig:linearCorrected}
\end{figure}

The true value of the two particle correlation function is shown in figure \ref{fig:linearCorrected} together with the value observed by the nominal detector and the corrected nominal detector.  The correction has removed the vast majority of the nominal detector effect, although a small residual detector effect remains.  Note that while the correction is quite large for the $\Delta\eta$ dependence, the same procedure applied to the $\Delta\phi$ dependence provides a much smaller correction that acts in the opposite direction; enhancing the peak at $\Delta\eta=0$ while slightly flattening the peak at $\Delta\phi=0$.  The bottom panels of figure \ref{fig:linearCorrected} show the difference between the true and corrected values of the two particle correlation function.
The correction to the nominal detector was also determined by using the alternative detector parameterisation to construct $X_{i}\left(1\right)...X_{i}\left(5\right)$ (whilst using the nominal parameterisation to construct the detector-level observable $X_{i}\left(0\right)$).  The corrected value thus obtained by using a mis-modelled detector to apply the correction is also shown in figure \ref{fig:linearCorrected}.  In this example, when using a polynomial fit of degree four to the detector iterations curve, the mis-modelled alternative detector in-fact shows a slightly better agreement between the corrected and true observables.  This better agreement occurs because the alternative parameterisation represents a stronger detector effect; a particle has an increased probability of being removed from the sample when using the alternative as opposed to nominal detector.  The alternative correction therefore over-corrects relative to the nominal correction, however this effect cancels with the slight under-correction provided by the nominal correction when using \emph{this particular} polynomial fit to the iterations curve.

\begin{figure}[h]
\begin{centreFigure}
\begin{minipage}[c]{0.49\linewidth}
\begin{overpic}[width=\columnwidth]{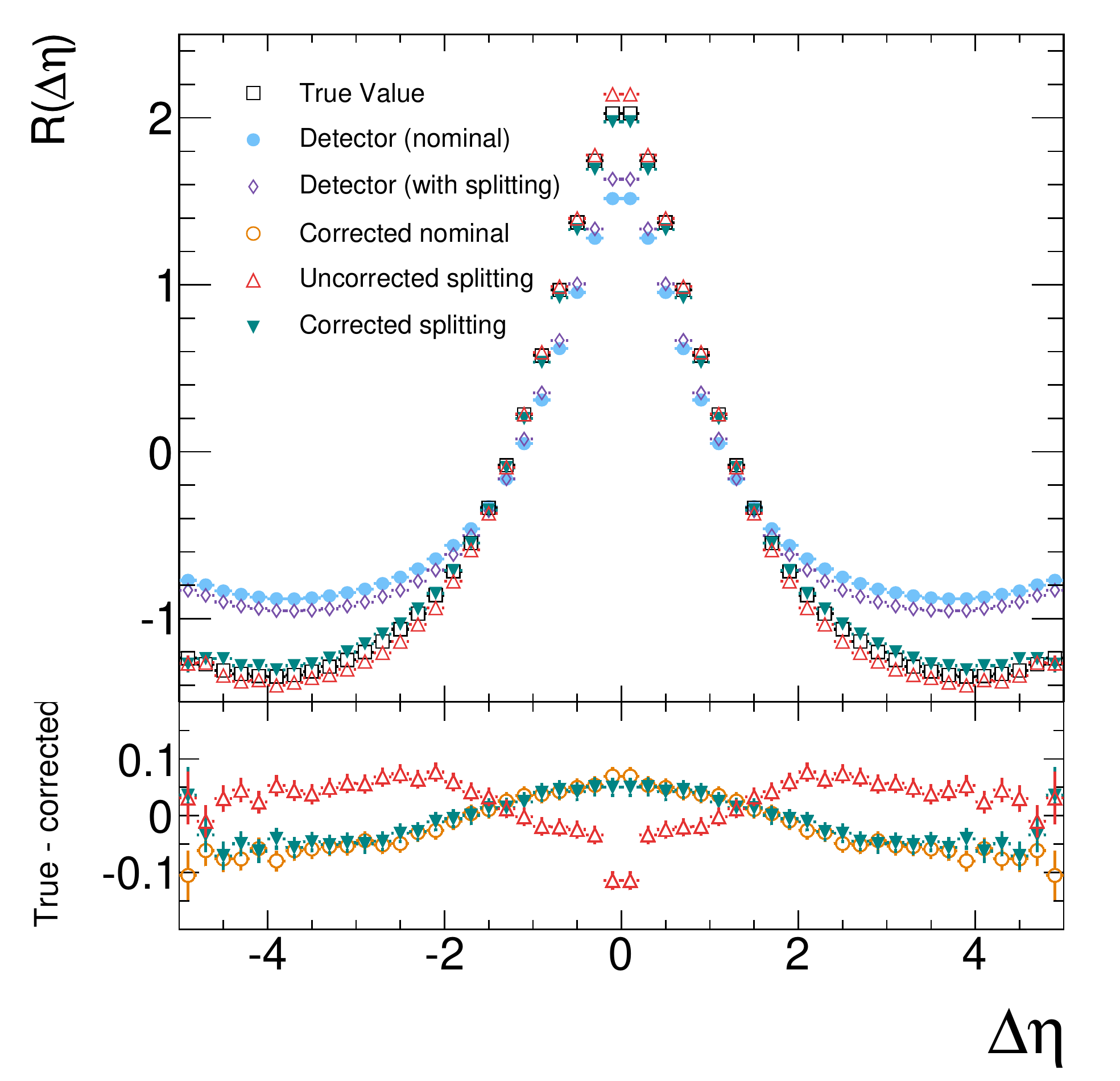}
\put(80,80){a)}
\end{overpic}
\end{minipage}
\begin{minipage}[c]{0.49\linewidth}
\begin{overpic}[width=\columnwidth]{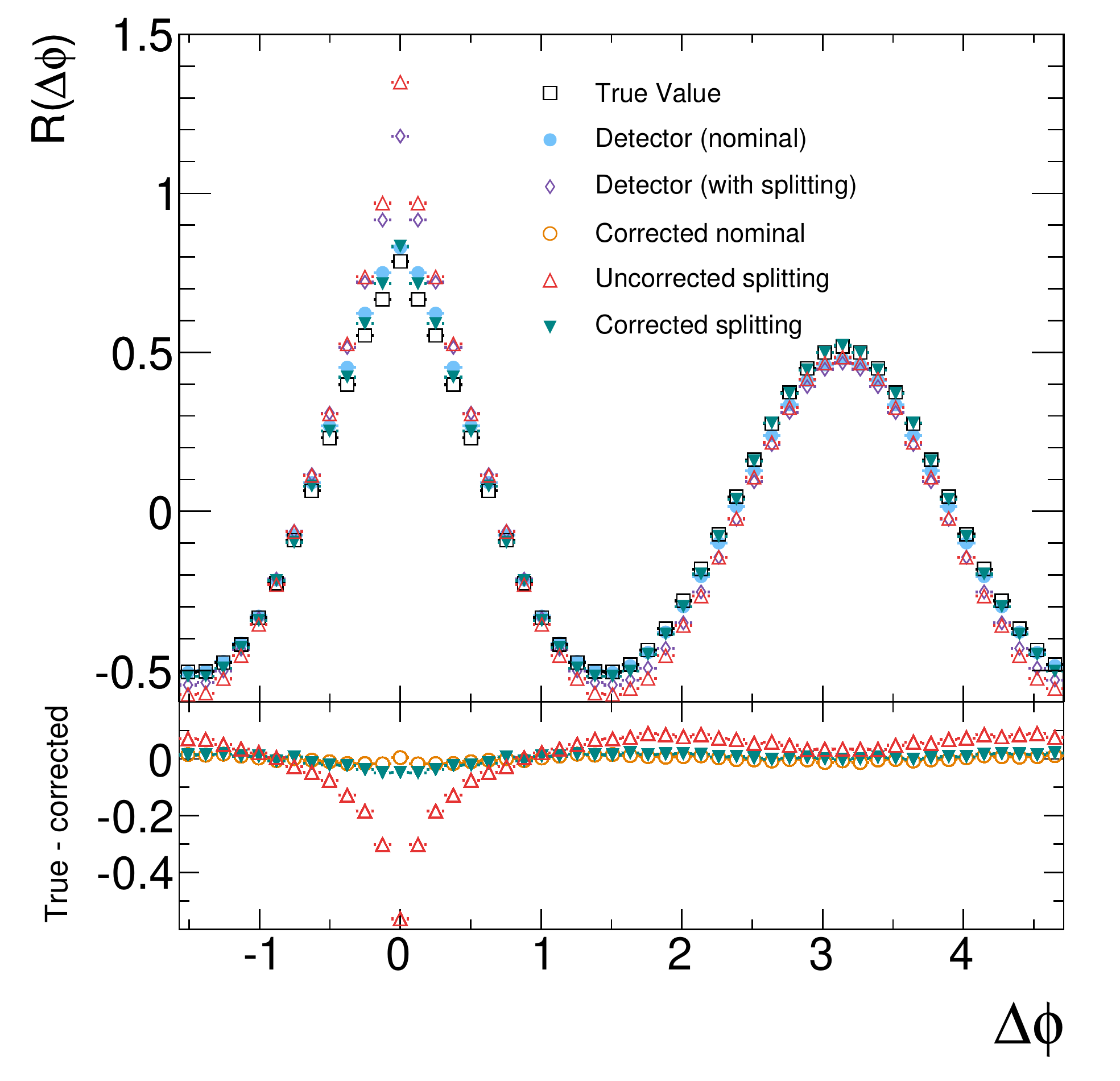}
\put(28,80){b)}
\end{overpic}
\end{minipage} 
\end{centreFigure}
\caption[Corrected and uncorrected two particle correlation function using a linear fit]{$\Delta\eta$ (a) and $\Delta\phi$ (b) dependence of the two particle correlation function corrected for the effects of particle splitting.  Corrections were applied using polynomial fits of degree four to the iterations curves (see figure \ref{fig:linearFits}). The nominal detector-level observable (without splitting) is shown as filled circles, while the additional effect of splitting is shown as un-filled diamonds.  Correcting the sample with splitting by using only the nominal detector parameterisation (un-filled triangles) can lead to a significant bias.  The effect of the splitting was completely removed (filled triangles) by performing the correction using a detector parameterisation that included the splitting effect.  The true value of the observable is shown as un-filled squares and the differences between the true and corrected values are shown in the bottom panel.}\label{fig:splitCorrected}
\end{figure}

The additional effects of particle splitting due to possible interactions with the detector material (equations \ref{eqn:splitProb} and  \ref{eqn:splitAngle}) are shown in figure \ref{fig:splitCorrected}.  Unlike the inefficiency for particle detection exhibited by the nominal detector, the addition of such secondary particles had a much stronger effect on the $\Delta\phi$ dependence of the correlation function than on the $\Delta\eta$ dependence.  In both cases the addition of particle splitting enhances the peak of the correlation function.  This effect is easily understood because the particles so-added are necessarily correlated.

Applying the the splitting effect to the $N=0$ detector level observable while using only the nominal detector to derive the correction led to a clear bias, as is shown in figure \ref{fig:splitCorrected}.  The main effect is that, after correction, the central peak at $\Delta\eta=0$ and $\Delta\phi=0$ is noticeably too high when using the nominal correction with the sample containing split particles.  This happens because the effect of splitting acts in the opposite direction to the effect of the loss of particles, therefore correcting \emph{only} for the loss of particles (as is the case with the nominal correction) leads to a correlation that is too strong.  

The splitting was corrected for by including the splitting parameterisation of equations \ref{eqn:splitProb} and \ref{eqn:splitAngle} in the detector function used to derive the $N=1...5$ iterations.  Fitting a polynomial to the resulting iterations curve and extrapolating to $N=-1$ corrected for both the tracks missing from the sample and the tracks introduced by the splitting.  The corrected two particle correlation function so obtained  shows an agreement with the true value that is as good as the corrected nominal sample, thus demonstrating that the effect of splitting can be entirely eliminated with this method.

\section{Variation of the Fits to the Detector Iterations}\label{sec:fits}

Fitting a polynomial of degree four produced a satisfactory correction that quite closely matches the generated hadron-level distributions.  In order to evaluate the success of the fit, other possible fitting functions were also explored.

\begin{figure}[h]
\begin{centreFigure}
\begin{minipage}[c]{0.49\linewidth}
\begin{overpic}[width=\columnwidth]{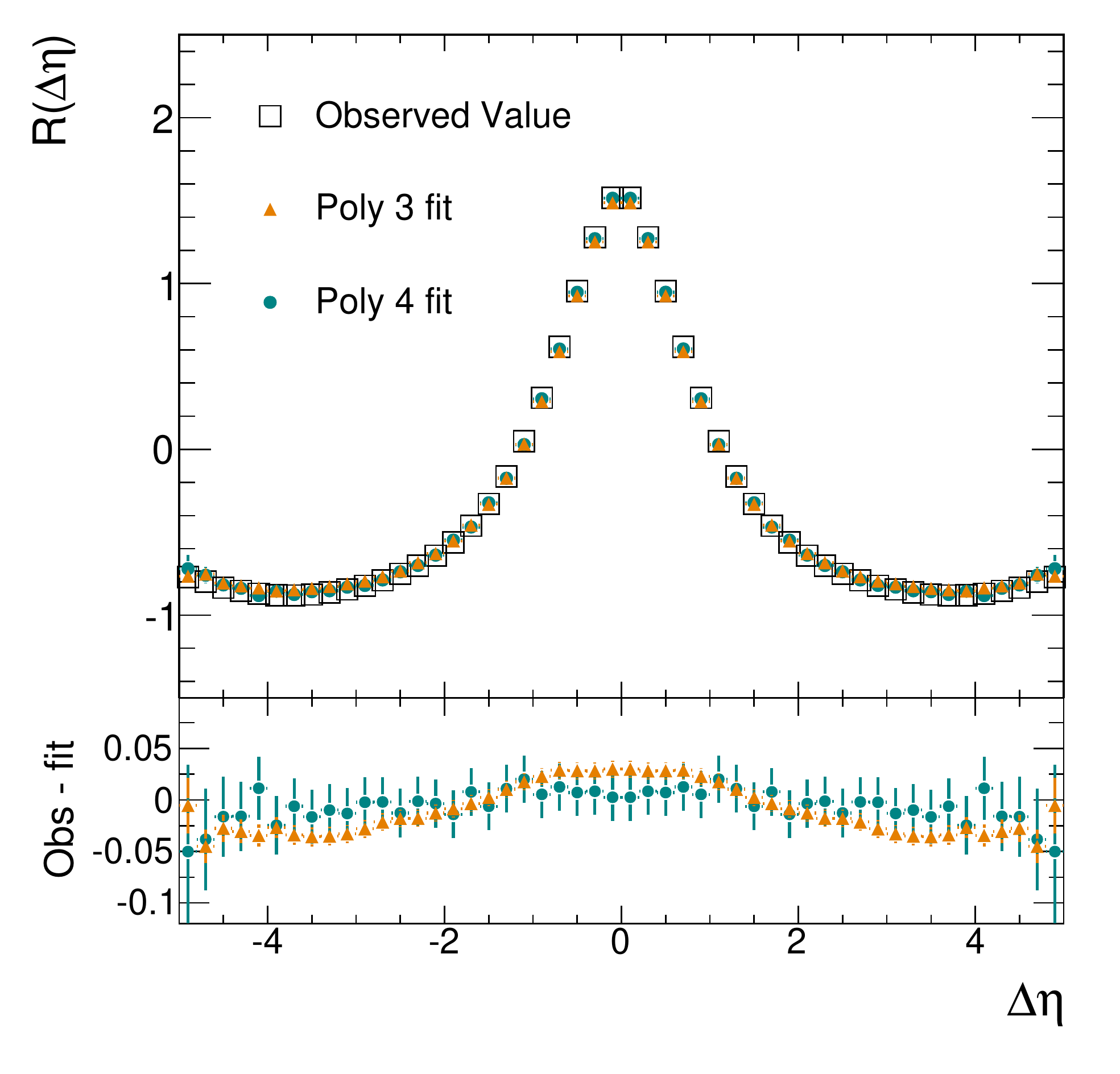}
\put(80,80){a)}
\end{overpic}
\end{minipage}
\begin{minipage}[c]{0.49\linewidth}
\begin{overpic}[width=\columnwidth]{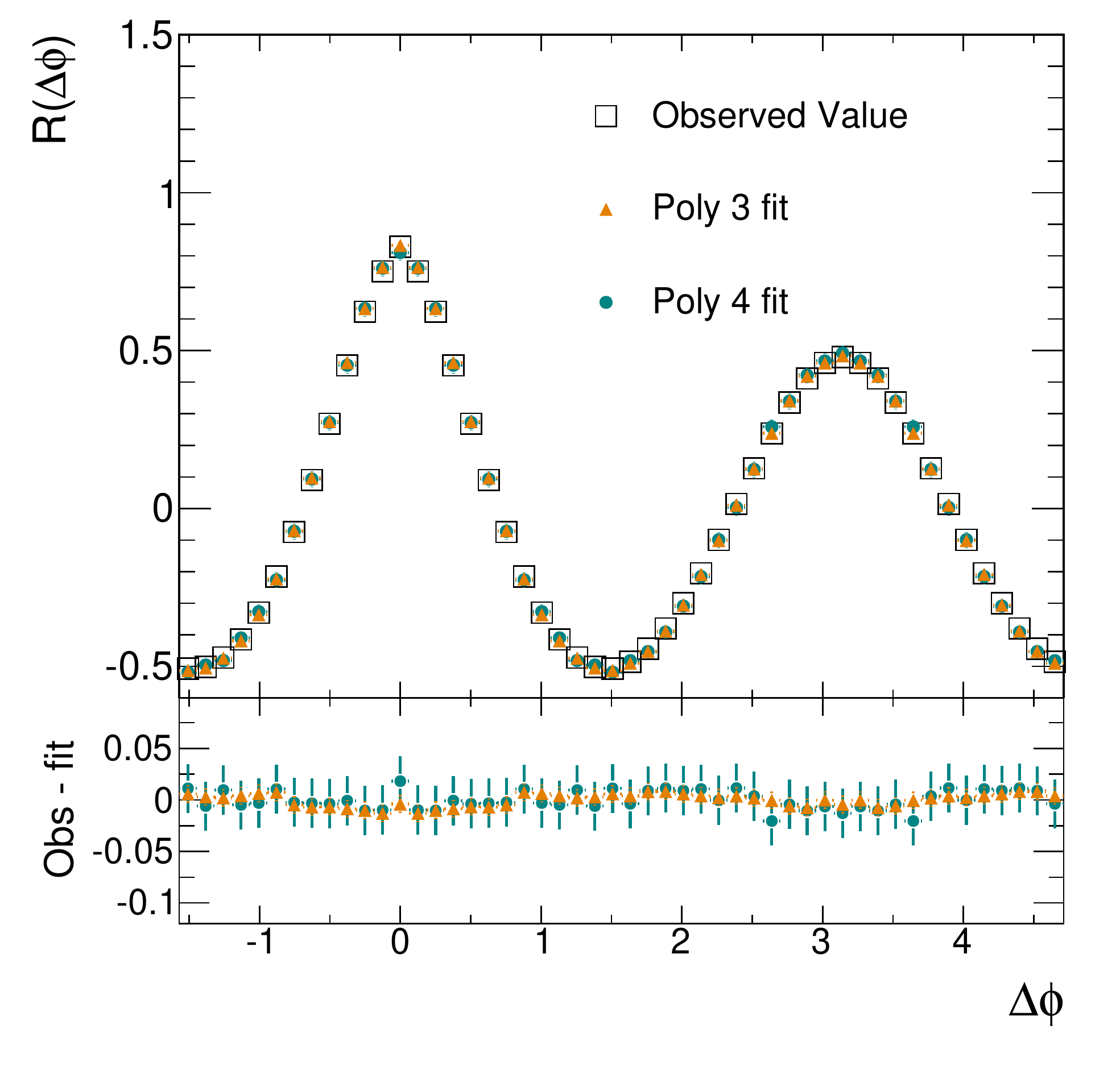}
\put(28,80){b)}
\end{overpic}
\end{minipage} 
\end{centreFigure}

\caption[]{The uncorrected ($N=0$) two particle correlation function as observed by the nominal parameterised detector compared to values of the uncorrected function that were estimated only from fits to further ($N \ge 1 $) applications  of the detector effect.  Fits to a polynomial of degree 3 (triangles) and degree 4 (circles) are shown for the $\Delta\eta$ (a) and $\Delta\phi$ (b) dependence of the two particle correlation function.}\label{fig:uncorrectedFit}
\end{figure}

As a test of the validity of the fitted function, polynomial fits of degree three and four were made to the points for which $N\geq1$ and extrapolated to the $N=0$ observed point.  Such a test would be possible during a real experiment, in which the true value of the observable would be unknown.  The resulting fits are shown in figure   \ref{fig:linearFits}, while the estimate of the $N=0$ observed uncorrected value of the correlation function is shown in figure \ref{fig:uncorrectedFit}. Figure   \ref{fig:uncorrectedFit} shows that, in this case, the fit using the polynomial of degree four generally exhibited less bias than the polynomial of degree three.  Note that while a polynomial of degree four was guaranteed to fit five data points more successfully than a polynomial of degree three, there was no guarantee that the extrapolation to $N=-1$ would be closer to the observed value.

\begin{figure}[h]
\begin{minipage}[c]{0.49\linewidth}
\begin{overpic}[width=\columnwidth]{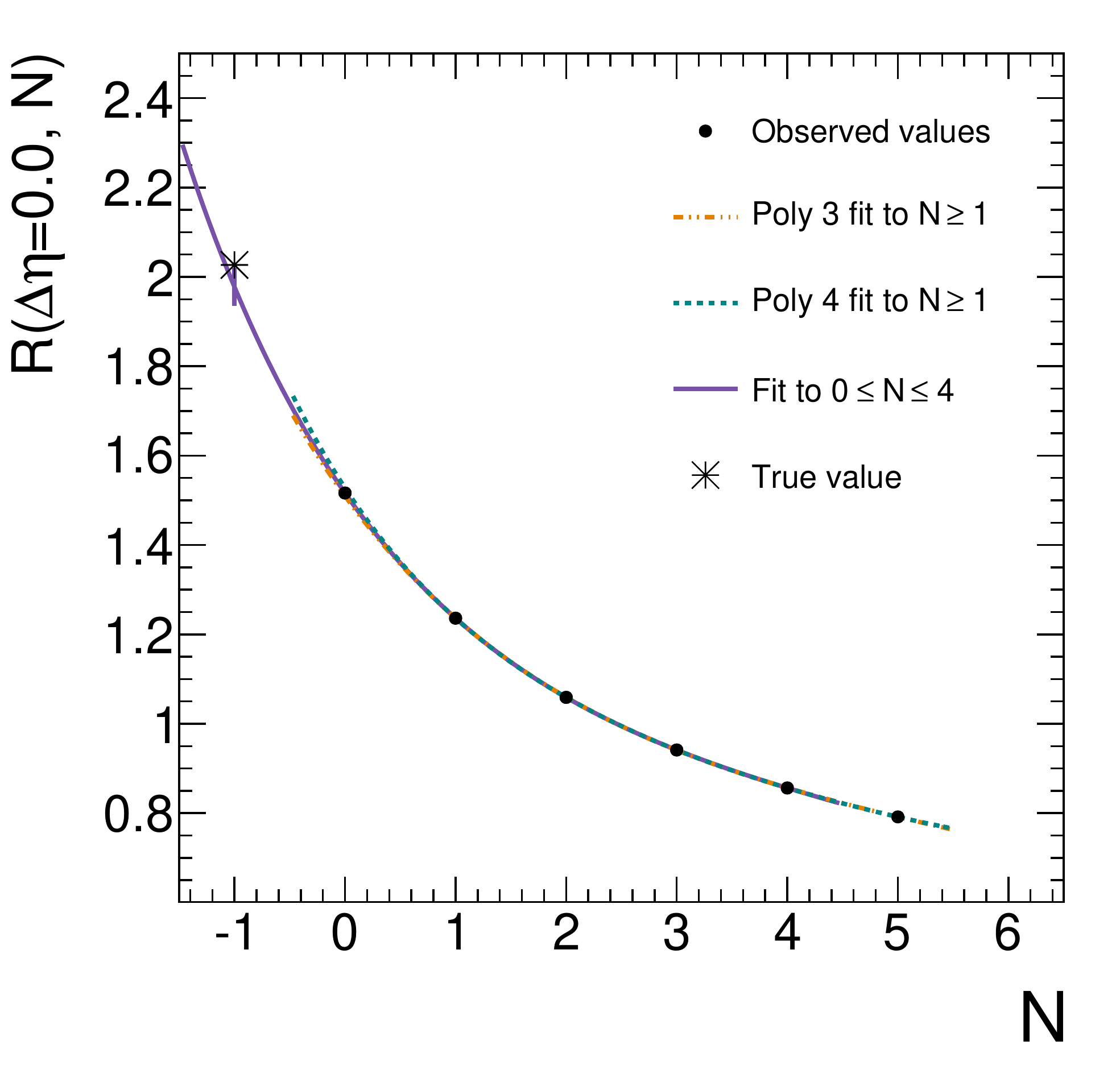}
\put(28,80){a)}
\end{overpic}
\end{minipage}
\begin{minipage}[c]{0.49\linewidth}
\begin{overpic}[width=\columnwidth]{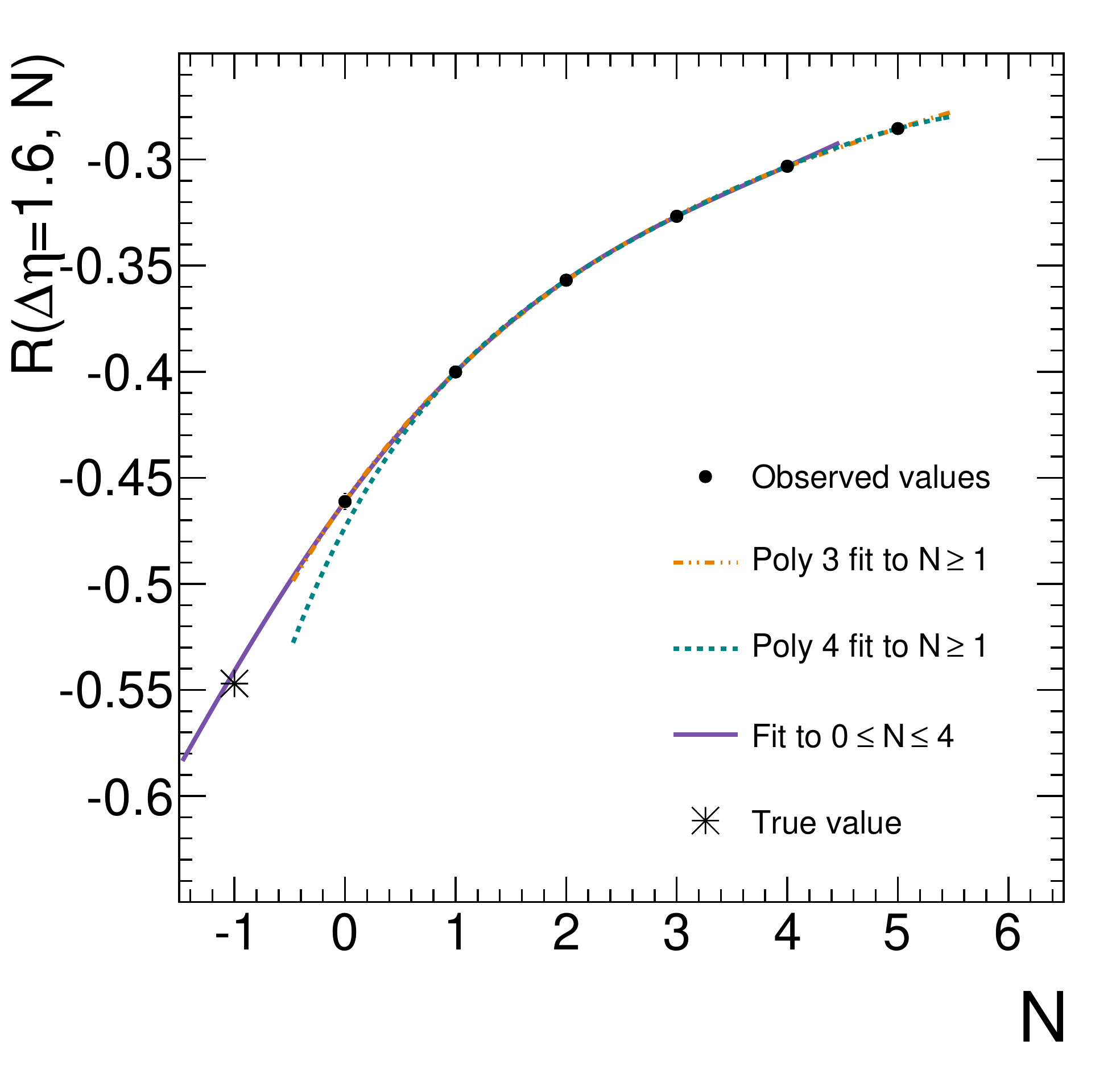}
\put(28,80){b)}
\end{overpic}
\end{minipage} 

\begin{minipage}[c]{0.49\linewidth}
\begin{overpic}[width=\columnwidth]{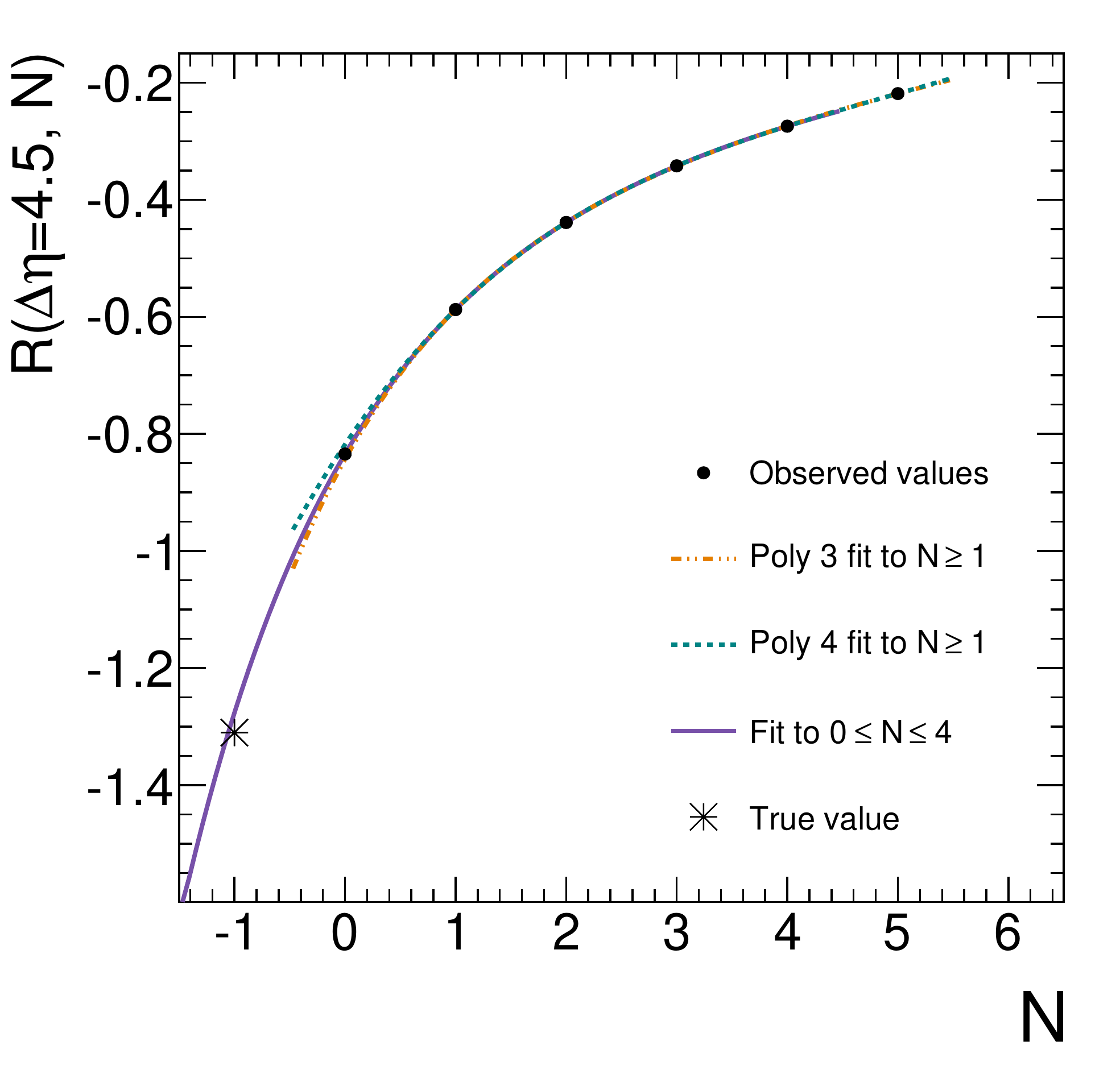}
\put(28,80){c)}
\end{overpic}
\end{minipage}\hspace*{0.05\linewidth}
\begin{minipage}[c]{0.53\linewidth}
\begin{flushleft}
\caption[Logarithmic fits to detector iterations curves]{Polynomial fits to the logarithm of the absolute value of the two particle correlation function Vs. the number of applications of the detector, $N$.  The value measured by the detector is at $N=0$, the true value is shown by a star at $N=-1$.   The observed $N=0$ value is estimated from both a cubic and quartic fit to the $N=1$ and higher points,  while the true $N=-1$ value is estimated from a quartic fit to the $N=0$ and higher points.  Fits are shown for three $\Delta\eta$ values of 0 (a), 1.6 (b) and 4.5 (c). }\label{fig:logFits}
\end{flushleft}
\end{minipage} 

\end{figure}

The iterations curve was very often (but not always) described reasonably well by an exponential decay because the number of particles removed at each detector application was proportional to the number of particles remaining in the sample.   Polynomial curves of degree three and four were therefore also fitted to the logarithm of the absolute value of each point in the iterations curve.  Note that, for a small number of points where the correlation function was near zero, logarithmic fits were not possible because after applying the detector effect several times, the correlation function can change sign.    When plotted in this way on a log scale where possible, the value of the observable was often close to linear in the number of detector applications.  The test of estimating the observed $N=0$ point from the $N\geq1$ points was performed and is shown in figure \ref{fig:logFits} for the same $\Delta\eta$ values as figure \ref{fig:linearFits}.  Fitting to the logarithm of the data points showed a generally good agreement between the estimated and true values, which was quite often better than the linear fit.  However, fitting to the logarithm of the data could, in some cases, lead to an unstable result due to large contributions from the higher order terms of the polynomial.  For such points it is better to use a fit without taking the logarithm of the data.

\begin{figure}[h]
\begin{centreFigure}
\begin{minipage}[c]{0.49\linewidth}
\begin{overpic}[width=\columnwidth]{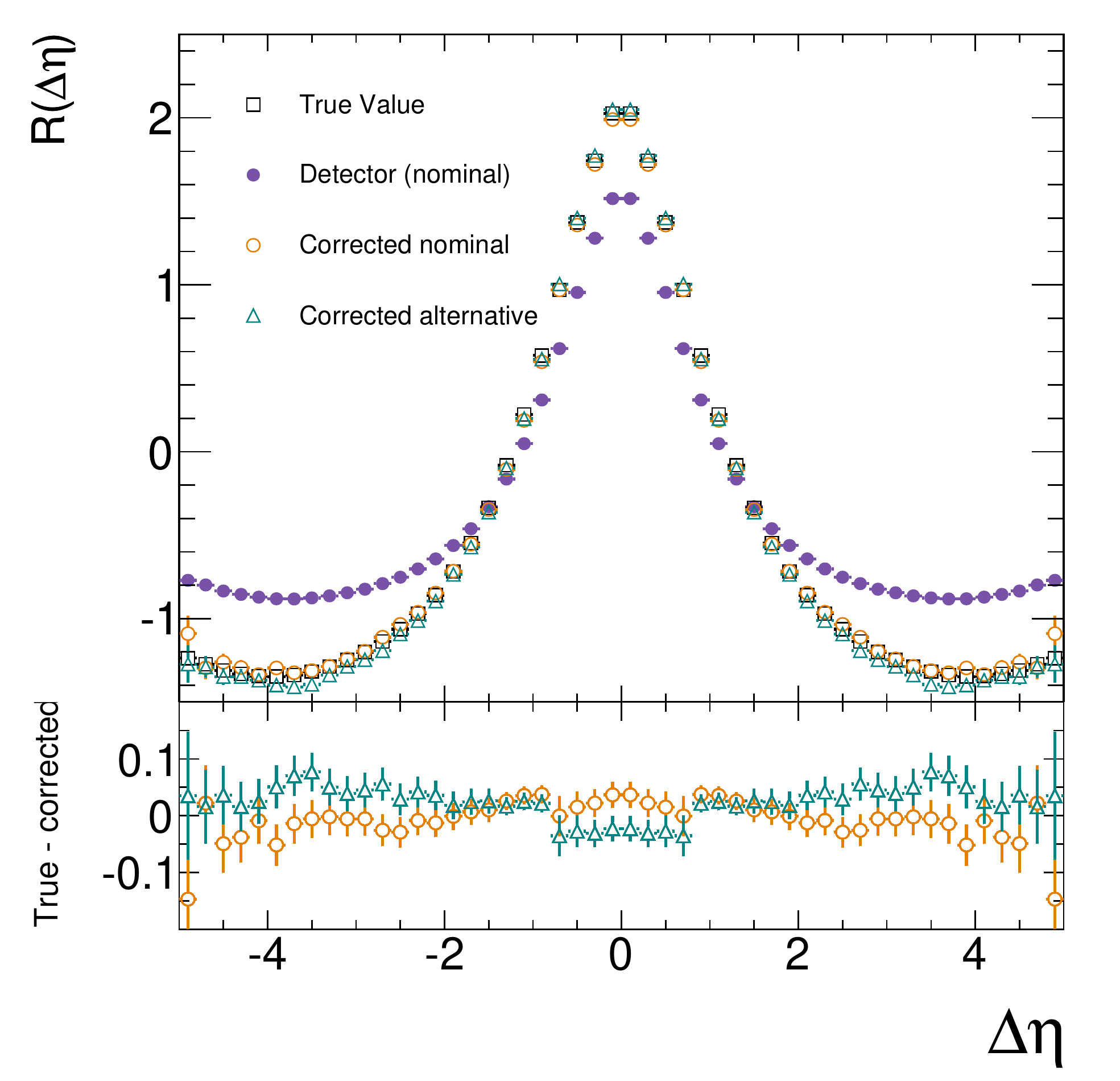}
\end{overpic}
\end{minipage}\hspace*{0.05\linewidth}
\begin{minipage}[c]{0.3\linewidth}
\begin{flushleft}
\caption[Corrected and uncorrected two particle correlation function using a logarithmic fit]{Corrected and uncorrected $\Delta\eta$ dependence of the two particle correlation function.  Corrections were applied using polynomial fits of degree four to the logarithm of the iterations curves (see figure \ref{fig:logFits}).  The $\Delta\eta$ curves are otherwise as described in figure \ref{fig:linearCorrected}.}\label{fig:logCorrected}
\end{flushleft}
\end{minipage} 
\end{centreFigure}

\end{figure}

The corrected value of  the correlation function using a logarithmic fit is shown in figure \ref{fig:logCorrected}.   The logarithmic fit resulted in corrected curves that exhibit somewhat less bias than the linear fits of figure \ref{fig:linearCorrected}, although the logarithmic fits tend to show slightly  larger uncertainties on the fit parameters.

\section{Discussion and Summary}

The HBOM method for correcting measurements for detector effects has the following advantages compared to other methods: 

\begin{itemize}
\item It is data-driven and model independent.  A reasonable approximation to the detector behaviour is sufficient.
\item It can be generalised to a large number of observables, including multi-dimensional observables.
\item Complicated observables, including those affected by correlations, may be corrected.
\item Histogram bin widths do not need to be optimised to avoid migrations.
\item No prior physics model is required.
\item The method is conceptually simple.
\end{itemize}

The method is ultimately limited by the approximation to and parameterisation of the detector response that is applied to the data.  In principle, a model of the detector may be constructed without reference to any MC simulation; however, in practice it is often the case that some sort of detector simulation is used to determine, for example, track-finding efficiencies.  If a detector response were derived from data alone then this method would be completely data-driven and independent of any simulation.  

Another limit on the method is the quality of the fits to the iterations curves and the level of confidence in those fits that can be obtained.  The fit encapsulates the behaviour of the detector; a bad fit to the iterations points will result in a bad description of the detector and a consequent poor correction.  For this note we used a heuristic approach to finding the fitting function, which recognises that no function is a-priori more correct than any other; we simply wanted a function that described the behaviour of the detector.  The performance of the fitted function can be estimated either by obtaining the original un-corrected data distributions from the additional applications of the detector effect or by performing a closure test on a simulated sample of events;  if the fit is good then the distributions obtained by passing hadron-level Monte Carlo events through a detector simulation and applying the correction procedure should agree with the original hadron-level results.  Testing the procedure on a given observable and experimental setup by performing such closure tests on an ensemble of different Monte Carlo event samples is, in general, a good test of the method.

A good correction also requires a reasonable description of the detector.  The variation shown here between the nominal and alternative detector parameterisations shows that the method has some degree of robustness against an imperfect parameterisation of the detector effect; however, as demonstrated by the effect of splitting, missing out an effect entirely can cause (or add to) a bias in the final distributions.  

The two particle correlation observable used here presents peaks in the distribution that are fixed in their $\Delta\eta$ and $\Delta\phi$ locations, regardless of the detector effect applied.  Other observables may, however, exhibit peaks whose position is dependent on the detector effect (particle multiplicity being a simple example).  If such moving peaks are present in the observable then fitting an iterations curve independently to each bin of the observable may very well not be the optimum corrections strategy because it ignores information about the overall change in the shape of the distribution.  In such cases it may be better to fit a function to the entire distribution and extrapolate the fit parameters to an $N=-1$ detector effect in order to obtain a correction factor.  We suggest that Bezier curves or their basis-spline generalisations may be ideal for this purpose; each additional application of the detector effect will cause the control points of the curve to migrate along a path.

While the example presented here shows a charged-track based measurement, there is nothing that would prevent this method being extended to measurements using calorimeter energy deposits.  In order to perform a correction on such an observable, a reasonable parameterisation of the energy and angular smearing produced in a calorimeter would be required.  Existing fast simulation packages may provide a suitable starting point for such parameterisations.  Correcting calorimeter observables in this way may even allow jet energies to be calibrated and corrected for soft phenomena such as multi-parton interactions.

\section*{Acknowledgements}

We thank Emily Nurse, Andrew Pilkington and Sharka Todorova for useful discussions and encouragement towards the HBOM method.  We would also like to give thanks and blame to Mario Campanelli for proposing the name ``HBOM.''   We are grateful to the STFC and CONACYT for funding this research in the UK and Mexico.

\end{document}